\documentclass[twocolumn,aps,amsmath,amssymb,prl,showpacs,preprintnumbers]
{revtex4}
\newcommand{\be}{\begin{equation}}
\newcommand{\ee}{\end{equation}}
\def\bq{\begin{eqnarray}}
\def\eq{\end{eqnarray}}
\def\beq{\begin{eqnarray}}
\def\eeq{\end{eqnarray}}
\def\ba{\begin{eqnarray}}
\def\ea{\end{eqnarray}}

\usepackage{dcolumn}
\usepackage{bm}

\newcommand{\M}{\ensuremath{{\cal M}}}

\begin{document}

\title{The structure of non-spacelike curves from a spacetime singularity}

\author{Pankaj S. Joshi}
\email{psj@tifr.res.in}

\affiliation{Tata Institute of Fundamental Research\\ Homi Bhabha Road\\ 
Mumbai 400 005, India}

\begin{abstract} We investigate here the causal structure of spacetime 
in the vicinity of a spacetime singularity. The particle and 
energy emission from such ultra-dense regions forming in gravitational 
collapse of a massive matter cloud is governed by the nature of 
non-spacelike paths near the same. These trajectories are examined 
to show that if a null geodesic comes out from the singularity, then 
there exist families of future-directed non-spacelike curves which 
also necessarily escape from the naked singularity. The existence of such 
families is crucial to the physical visibility of the singularity. 
We do not assume any underlying symmetries for the spacetime, and 
earlier considerations on the nature of causal trajectories emerging 
from a naked singularity are generalized and clarified.

\end{abstract}

\maketitle

One of the most important questions in gravity physics today
is whether spacetime singularities could be visible to faraway
observers in the universe. The occurrence of singularities is a generic 
feature of Einstein's theory of gravity, as shown by the singularity 
theorems in general relativity
\cite{HE}. 
These could occur in cosmology at the origin of the universe, 
or in the continual gravitational collapse of a massive star at the 
end of its life cycle.

Spacetime singularities are the regions where the physical 
conditions such as densities and curvatures are at their extreme. 
While the big bang singularity of cosmology is visible in principle, 
which gave rise to the universe as a whole, we cannot actually see it. 
On the other hand, when a massive star collapses continually 
under gravity, the eventual spacetime singularity can be either 
hidden within an event horizon of a black hole, or it could be 
visible to outside observers depending on how the collapse of the 
cloud dynamically evolves and the causal structure in the vicinity
of the singularity. The visibility or otherwise of singularities 
in gravitational collapse is an issue basic to the foundations of 
black hole physics. A visible naked singularity forming in collapse 
could provide an opportunity for the physical effects taking place 
in these extreme gravity regions to be observable to outside 
observers in the universe.

It is some times argued that extreme physical conditions
prevail closer to spacetime singularities, where quantum effects must
be important and must be incorporated into considerations.
This is plausible and in the very late stages of gravitational collapse 
quantum effects have to be taken into account, which could possibly 
resolve a classical naked singularity (see e.g.
\cite{GJS}).
The point, however, is that in such a case also the causal
structure of spacetime in the vicinity of the super-ultradense 
regions forming in collapse would be such that these would be
visible in principle to an external observer. In other words,
the quantum effects taking place in the regions with arbitrarily
high matter densities and curvatures could be seen by the 
external observers. Then, even though the naked singularity 
itself may be resolved, this provides an interesting physical 
scenario to possibly observe quantum gravity effects. The communication 
of physical effects from such extreme regions would be again 
governed by the existence of families of non-spacelike paths from 
the vicinity of the singularity. It is therefore important to 
understand the structure of such families within a gravitational 
collapse framework.

The formation and existence of naked singularity in a 
gravitationally collapsing matter cloud has been extensively
investigated in past decade or so (see e.g. 
\cite{rev} 
for some reviews, and references therein). This is typically deduced 
based on the causal structure of spacetime and the nature of trapped 
region in the vicinity of the singularity. Let the continual gravitational 
collapse develop from a regular initial data from an initial 
spacelike surface, and result into the development of a spacetime 
singularity. If there is an outgoing future directed null geodesic, 
which terminates in the past at the singularity, then the singularity
is at least locally visible. Such a locally naked singularity will   
be globally visible provided the outgoing null geodesic would come out 
of the boundary of the collapsing cloud to reach a faraway observer 
in the spacetime. On the other hand, if the collapse terminates into 
a black hole final state, no non-spacelike curves would escape 
from the singularity. It is typically seen that the nature of 
initial density and pressure profiles and the dynamical evolutions 
as allowed by the Einstein equations actually determine the final 
state of the cloud in terms of either a black hole or a naked 
singularity.

An important physical issue then would be whether such a naked 
singularity forming in gravitational collapse could radiate away 
energy and particles. The physical visibility, or otherwise, of the 
singularity developing towards the end of continual collapse of the 
matter cloud depends crucially on the existence and structure of 
families of non-spacelike trajectories coming out from its vicinity. 
For example, if only a single photon could escape from the singularity, 
it will be non-visible to an external observer for all practical 
purposes. On the other hand, existence of families of future directed 
non-spacelike paths could make the singularity visible for outside 
observers. Also, any material particles would escape from the 
vicinity of the singularity only if there are timelike curves 
escaping away from these ultra-dense regions.

As pointed out above, a standard technique to deduce the 
development of a naked singularity in gravitational collapse has 
been to examine the existence of a future directed outgoing radial 
null geodesic from the same. Radial null geodesics are, however, 
somewhat special paths. The actual physical appearance of these 
ultra-dense regions will be typically determined by the non-radial null 
trajectories from the same, and the energy emission, if any will be governed 
by the timelike curves and other non-spacelike trajectories escaping 
away from the singularity. For this reason, several authors have 
considered the possibility of non-radial null geodesics coming out from 
a naked singularity in the context of spherically symmetric dust 
collapse models
\cite{dust}. 
Also, families of non-spacelike and timelike geodesics have been 
worked out in the self-similar perfect fluid collapse,
and the Vaidya radiation collapse models
\cite{JD}.
Most of these considerations have been in the framework of spherically 
symmetric spacetimes, at times together with other symmetry conditions imposed 
such as self-similarity of the models, and within the framework of a 
specific matter model.

A general consideration, however, on the nature of non-spacelike 
trajectories near a naked singularity will be of interest from such 
a perspective. We examine here the non-spacelike trajectories from a 
naked singularity in a general manner, and show that if a radial null 
geodesic comes out, then large families of non-spacelike curves 
also necessarily come out from the singularity. In other words, it 
is seen that the existence of a radial null geodesic is sufficient to 
ensure the existence of families of timelike and non-spacelike trajectories 
escaping, and in this sense a single photon escaping in a radial 
direction from the singularity is never an isolated phenomenon. This 
generalizes and clarifies earlier considerations in this direction
such as 
\cite{dust}, 
without assuming any symmetry conditions on the underlying spacetime 
or any specific matter model for the collapse. As we point out below,
the naked singularity generates an indecomposable future set, the
null boundary of which gives all the null generators, including the
null geodesics, coming out from the singularity.

Let us consider a continual gravitational collapse which
ends in a naked singularity, that is, the causal structure near the 
singularity is such that a null geodesic trajectory $\gamma$ 
comes out from the same. Specifically, $\gamma$ is 
future-directed, which in the past terminates at the singularity, 
and is therefore a past-incomplete null geodesic. To examine in
general the possible existence and nature of non-spacelike curves 
coming out of this naked singularity, we use here the causal 
boundary construction as developed by Geroch, Kronheimer and Penrose 
\cite{geroch}.
(Our notations are same as given by
\cite{HE}
and the spacetime $\M$ is taken to satisfy a suitable causality
condition such as strong causality, which rules out existence
of closed timelike curves.)
In this procedure, a boundary is attached to the regular 
spacetime manifold, which includes spacetime singularities as 
well as the points at infinity
\cite{rem}. 
An open set $W$ in the spacetime is 
called a {\it future set} if it contains its own future,
that is, we 
have $I^{+}(W) \subset W$. Further, a future set $W$ is called an {\it 
indecomposable future set (IF)} if it cannot be expressed as the union 
of two proper subsets which are themselves future sets. Indecomposable
past sets ($IP$s) are similarly defined. The idea of the causal 
boundary construction is to divide the collection of $IF$s and $IP$s 
into two classes, namely the one representing regular points of 
the spacetime, and the other class giving all its boundary points or 
the ideal points, which include spacetime singularities as well as
points at infinity. The collection of $IF$s (or $IP$s) can be divided 
into the two parts as follows. For a set $W$ which is an $IF$, 
if there exists 
an event in the spacetime $p\in \M$ such that $W=I^{+}(p)$, then 
$W$ is called a {\it proper IF} or a $PIF$. All other $IF$s are 
called {\it terminal IFs} or $TIF$s, which represent spacetime 
singularities and the points at infinity.

Consider now the set $I^{+}(\gamma)$ where $\gamma$ is 
any null geodesic curve coming out of the singularity. It is 
then a future set, because for any $p \in I^{+}(\gamma)$, 
we have $I^{+}(p) \subset I^{+}(\gamma)$. We can see that 
$I^{+}(\gamma)$ is an indecomposable future set, or an $IF$. 
To show this, we use a somewhat modified version of 
the proof of Theorem 2.1 of
\cite{geroch}.
Suppose $I^{+}(\gamma) = A \cup B$ with $A$ and $B$ both 
being future sets. If neither $A$ is fully contained in $B$ or 
vice-versa, we can then find two events $x, y$ such that 
$x \in A-B$ and 
$y \in B-A$.  We have both $x,y \in I^{+}(\gamma)$ so there 
are points $x',y' \in \gamma$ such that $x \in  I^{+}(x')$ and
$y \in  I^{+}(y')$. But $x'$ and $y'$ are causally related so
suppose now that $x'$ is in the past of $y'$ on $\gamma$. 
Then there is a null geodesic from $x'$ to $y'$, and there is a 
timelike
curve from $y'$ to $y$ as above. This implies that there must
be a timelike curve from $x'$ to $y$ (see e.g.
\cite{HE}, 
p.183). It follows that $y \in I^+(x')$. As we already have
$x \in I^+(x')$, this implies that $x, y \in I^+(x')$.
Therefore, $x'$ lies in the intersection of the sets $I^-(x)$
and $I^-(y)$, which is an open set and so contains a neighbourhood $N$
of $x'$. Let $z$ be an event in $I^+(x') \cap N$, then 
$z \in I^+(\gamma)$ and so has to be in one of the future sets 
$A$ or $B$. Suppose it is in $A$, then since there are future 
directed timelike curves from $z$ to both $x$ and $y$, it follows
that both $x, y \in A$, which is a contradiction.
Hence it follows that $I^{+}(\gamma)$ has to be an $IF$. Since $\gamma$ 
is a past-incomplete null geodesic, there is no regular point 
$p \in \M$ such that $ I^{+}(p) =  I^{+}(\gamma)$, and hence  
it follows that $I^{+}(\gamma)$ is necessarily a $TIF$.

The $TIF$ set $I^{+}(\gamma)$ here represents a boundary 
point of the spacetime which is the naked singularity. While 
the naked singularity formation as endstate of a continual 
gravitational collapse has been investigated extensively in past 
decade or so (especially 
within the framework of spherically symmetric collapse and for 
certain non-spherical examples), the main technique there has 
been to show that there exists a radial null geodesic coming out 
in the future and which terminates in the past at the 
spacetime singularity 
\cite{rev} 
It is thus seen that the gravitational collapse from 
regular initial matter profiles could result into either of the 
black hole or naked singularity endstates, depending on the nature of 
the initial data from which it evolves and the dynamical evolutions 
of the collapsing cloud as allowed by the Einstein equations.

However, as remarked above, if one is to examine 
the visibility and other related physical characteristics 
of a naked singularity that formed in gravitational collapse, 
then it is important to examine and understand the structure 
of families of non-spacelike curves from 
the same.  Again, if non-spacelike curves come out of the singularity 
but do not go out of the boundary of the collapsing cloud then the 
singularity will be only locally visible but the outside 
observers would not be able to see the same. It is necessary 
therefore to understand the structure of non-spacelike curves 
from a naked singularity in general.

It is now possible to do this as below.
Since the set $S = I^{+}(\gamma)$ is a $TIF$, 
it follows from 
\cite{geroch}
that in this case there must exist a past-inextendible 
timelike curve $\lambda$ such that $S = I^{+}(\lambda)$.  In the 
case of collapse ending in a naked singularity and a radial 
null geodesic $\gamma$ escaping from the same, we thus see 
that the set $I^{+}(\gamma)$ {\it is} a $TIF$, and so 
by the above result, there is a  timelike curve $\lambda$ generating
this $TIF$, in the sense that $S = I^{+}(\lambda)$. Since both the 
non-spacelike trajectories $\gamma$ and $\lambda$ represent the 
same ideal or boundary point of the spacetime which is the naked singularity, 
and since $I^{+}(\gamma) =  I^{+}(\lambda)$ by definition, it follows that
the future-directed timelike curve $\lambda$ must terminate in the 
past at the naked singularity. In other words, we have shown
that there exists is a timelike curve $\lambda$ which escapes away
to future and which terminates in the past at the naked singularity.

It follows that if $p \in \lambda$ and $q$ is any 
other event such that $q \in I^{+}(p)$, then there are 
timelike curves from the naked singularity to $q$. This proves the 
existence of families of infinitely many future-directed non-spacelike 
trajectories escaping away from the naked singularity.   
In general, if $\lambda'$ is any other future directed 
non-spacelike curve such that
$I^{+}(\lambda') = I^{+}(\lambda)$, then it follows that they
all represent the same $TIF$, which is the naked singularity, and
that $\lambda'$ terminates in the past at this singularity.
Thus there is an infinity of future going non-spacelike curves
which emerge from the singularity.

We thus see that the usual method employed to show 
the existence of a naked singularity in collapse, which establishes 
the existence of a radial null geodesic escaping away from the 
same, is sufficient to lead to the existence of infinite families of 
future going non-spacelike curves from the naked singularity 
as seen here. In the present consideration, we no longer need any 
special symmetry assumptions on the spacetime such as spherical symmetry, 
self-similarity etc, or any specific form of matter model such 
as dust equation of state which are usually assumed in such 
discussions.

In particular, this also clarifies and generalizes the earlier 
results on dust collapse and other models mentioned above, which have 
particularly focussed on non-spacelike null geodesics. The null 
geodesics of the spacetime have of course a special role to play as 
far as the visibility of the singularity is concerned. From such a 
perspective, let us briefly discuss the existence of radial versus the 
non-radial families of null geodesics from a naked singularity. 
Suppose a radial null geodesic comes out from the naked singularity $S$ 
developing in a continual collapse. In that case, as seen above, 
there exists a timelike curve $\lambda$ generating the $TIF$ 
set $I^{+}(\lambda)$ which represents the boundary point $S$. All other 
future-directed non-spacelike curves $\gamma$ which 
satisfy  $I^{+}(\lambda) = I^+(\gamma)$ generate the same $TIF$ 
representing the boundary point $S$, and they give the families
of particle or photon trajectories escaping away from the naked
singularity. The boundary of this future set which is a $TIF$, is
a three-dimensional null hypersurface which is ruled by the radial as well 
as non-radial null geodesics generators $\gamma$s, which are all 
incomplete when extended in the past, and which all have the property 
that  $I^{+}(\lambda) =  I^{+}(\gamma)$. This shows that the existence 
of a radial null geodesic is sufficient to give families of non-radial 
null geodesics as well, coming out from the singularity. This generalizes 
the earlier results on existence of non-radial null geodesics from 
the singularity for the spherically symmetric dust collapse, when a 
radial null geodesic came out from the same.

It is thus seen that once a singularity is naked, it gives
rise to infinitely many null as well as timelike curves to escape
away from the same. In this sense, the emission of paths representing 
particle or photon trajectories from the visible singularity is a generic 
phenomena. This is a necessary condition for the singularity to give 
rise to any physical effects which may possibly be observed by 
external observers. We have not discussed in the present consideration  
the global visibility of the singularity, that is, once the families
of non-spacelike curves come out of the naked singularity when they 
will actually cross the boundary of the cloud to escape to an outside 
observer. It is known, however, in several cases including spherical 
dust collapse, that whenever a singularity is locally naked then 
one can choose rest of the free functions in the model so as to make 
it globally visible
\cite{remark}.   
A discussion on global visibility in a more general context 
of perfect fluids is recently given by
\cite{giambo}.


\begin{thebibliography}{99}
\bibitem{rem} A boundary attachment to the spacetime manifold is
essential to treat the regular spacetime events together with its
singularities and points at infinity in a unified manner. There are
different ways to attach a boundary to the spacetime, and they all
do not necessarily give the same result. We have used here 
the approach as given in
\cite{geroch} 
as it depends basically only on the causal 
structure of spacetime, which is more fundamental as compared to, 
for example, the differential structure of the spacetime manifold.  
Also, from a physical point of view, each ideal point here is
directly associated with the region of spacetime which it can 
influence, or which it would be influenced by.
\bibitem{remark} It is known, e.g. in the case of dust collapse,
that once the singularity is locally naked, the choice of a suitable
behaviour of the mass function (which is a free function, subject 
only to some physical conditions such as an energy condition and 
regularity of the initial data) away from the center, allows 
the null rays to come out from the boundary of the cloud (see
\cite{JD}). 
It may also be noted that in certain classes of self-similar 
collapse, once the singularity is locally naked it becomes
necessarily globally visible. In any case, as there is no scale 
in the problem, once the singularity is locally visible, an
observer within a large enough black hole will still be able to
see it for a long enough time. In such a scenario, the escape
of rays outside the boundary of the cloud would not 
be crucial.
\end{thebibliography}
\end{document}